\documentclass[sigplan, screen]{acmart}
\AtBeginDocument{%
  }

\setcopyright{acmlicensed}
\copyrightyear{2026}
\acmYear{2026}
\acmDOI{XXXXXXX.XXXXXXX}
\acmConference[Conference acronym 'XX]{Make sure to enter the correct
  conference title from your rights confirmation email}{June 03--05,
  2018}{Woodstock, NY}

\acmISBN{978-1-4503-XXXX-X/2018/06}

\usepackage{enumitem}
\usepackage{subcaption}
\usepackage{makecell} % 在导言区加入此包，用于单元格内换行
\usepackage[absolute,overlay]{textpos} % 提供 textblock* 环境

\begin{document}

% GPR logo
\begin{textblock*}{\paperwidth}(0.5cm, 0.5cm)
  \includegraphics[height=1.3cm]{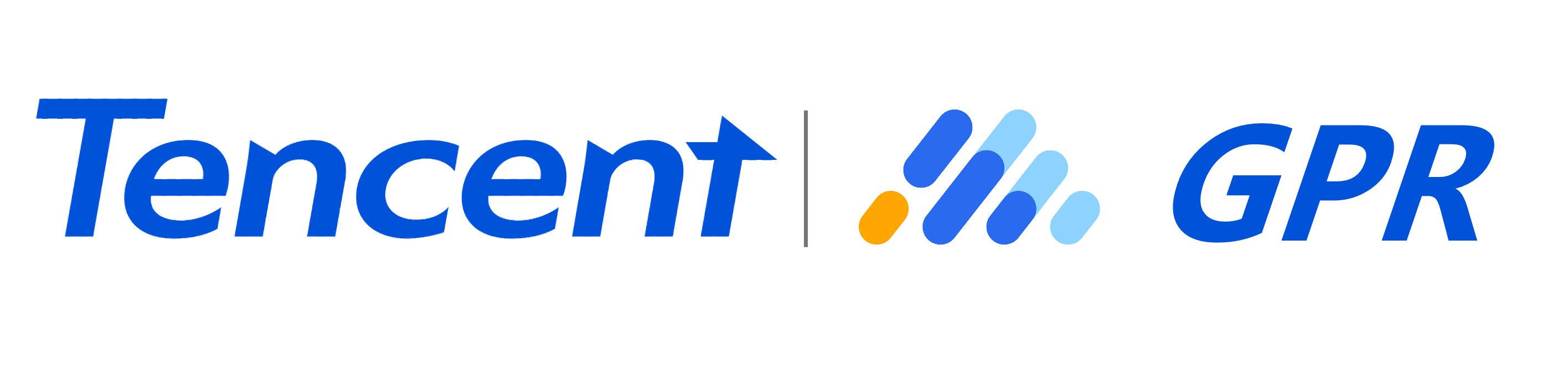}
\end{textblock*}

\title[OneRanker: Unified Generation and Ranking with One Model]{OneRanker: Unified Generation and Ranking with One Model in Industrial Advertising Recommendation}

\author{Dekai Sun}
\authornote{Equal contribution.}
\affiliation{%
  \institution{Tencent Inc.}
  \city{Beijing}
  \country{China}
}
\email{dekaisun@tencent.com}

\author{Yiming Liu}
\authornotemark[1]
\affiliation{%
  \institution{Tencent Inc.}
  \city{Beijing}
  \country{China}}
\email{lucienymliu@tencent.com}

\author{Jiafan Zhou}
\authornotemark[1]
\affiliation{%
  \institution{Tencent Inc.}
  \city{Beijing}
  \country{China}
}
\email{jiafanzhou@tencent.com}

\author{Xun Liu}
\authornotemark[1]
\affiliation{%
 \institution{Tencent Inc.}
  \city{Beijing}
  \country{China}}
\email{reubenliu@tencent.com}

\author{Chenchen Yu}
\affiliation{%
  \institution{Tencent Inc.}
  \city{Beijing}
  \country{China}}
\email{chenchenyu@tencent.com}

\author{Yi Li}
\affiliation{%
  \institution{Tencent Inc.}
  \city{Beijing}
  \country{China}}
\email{sincereli@tencent.com}

\author{Jun Zhang}
\authornote{Corresponding author.}
\affiliation{%
  \institution{Tencent Inc.}
  \city{Beijing}
  \country{China}}
\email{neoxzhang@tencent.com}

\author{Huan Yu}
\affiliation{%
  \institution{Tencent Inc.}
  \city{Beijing}
  \country{China}}
\email{huanyu@tencent.com}

\author{Jie Jiang}
\affiliation{%
  \institution{Tencent Inc.}
  \city{Beijing}
  \country{China}}
\email{zeus@tencent.com}

\renewcommand{\shortauthors}{Sun et al.}

\begin{abstract}
The end-to-end generative paradigm is increasingly adopted in advertising recommendation systems, driving a shift from traditional cascaded architectures towards unified modeling. However, practical deployment faces three core challenges: the misalignment between interest objectives and business value, the target-agnostic limitation of generative processes, and the disconnection between generation and ranking stages. Existing solutions often fall into a dilemma where single-stage fusion induces optimization tension, while stage decoupling causes information loss. To address this, we propose \textbf{OneRanker}, achieving architectural-level integration of generation and ranking. First, we design a value-aware multi-task decoupling architecture. By leveraging task token sequences and causal mask, we separate interest coverage and value optimization spaces within shared representations, effectively alleviating target conflicts. Second, we construct a coarse-to-fine collaborative target awareness mechanism, utilizing Fake Item Tokens for implicit awareness during generation and a ranking decoder for explicit value alignment at the candidate level. Finally, we propose input-output dual-side consistency guarantees. Through Key/Value pass-through mechanisms and Distribution Consistency (DC) Constraint Loss, we achieve end-to-end collaborative optimization between generation and ranking. The full deployment on Tencent's WeiXin Channels advertising system has delivered consistent improvements on key business metrics (e.g., GMV-Normal +1.34\% in the initial rollout phase), offering a practical and deployable design for generative advertising recommendation.
\end{abstract}

\begin{CCSXML}
<ccs2012>
   <concept>
       <concept_id>10002951.10003317.10003338.10003343</concept_id>
       <concept_desc>Information systems~Learning to rank</concept_desc>
       <concept_significance>500</concept_significance>
       </concept>
   <concept>
       <concept_id>10002951.10003317.10003347.10003350</concept_id>
       <concept_desc>Information systems~Recommender systems</concept_desc>
       <concept_significance>500</concept_significance>
       </concept>
   <concept>
       <concept_id>10002951.10003227.10003447</concept_id>
       <concept_desc>Information systems~Computational advertising</concept_desc>
       <concept_significance>500</concept_significance>
       </concept>
 </ccs2012>
\end{CCSXML}

\ccsdesc[500]{Information systems~Learning to rank}
\ccsdesc[500]{Information systems~Recommender systems}
\ccsdesc[500]{Information systems~Computational advertising}

\keywords{Recommender Systems, Generative Modeling, Advertising}

\maketitle

\section{Introduction}
\label{sec:introduction}

Advertising and recommendation systems in large-scale internet applications are undergoing a fundamental transformation from traditional cascaded architectures to end-to-end generative paradigms ~\cite{onerec,gpr,dai2025onepiece,guo2025onesug,huang2025towards}. Driven by breakthroughs in large language models (LLMs) ~\cite{achiam2023gpt,grpo} and sequence modeling techniques~\cite{zhang2025onetrans,chai2025longer}, this shift has fundamentally redefined how advertising platforms match users with relevant commercial content. 
As demonstrated by industrial systems such as the OneRec~\cite{onerec,onerecv2_2024} series and GPR~\cite{gpr}, generative recommendation has emerged as a promising approach that formulates recommendation as an autoregressive generation task based on hierarchical semantic IDs, enabling direct optimization of recommendation objectives. 
However, as generative methods penetrate deeper into the ranking stage of advertising systems, three critical challenges have become central to the next-generation architecture design: unified modeling of interest objectives and business value within the generation process, precision degradation caused by target-agnostic generation, and representation disconnection between generation and ranking.

\begin{figure}[t]
\centering
\includegraphics[width=0.45\textwidth]{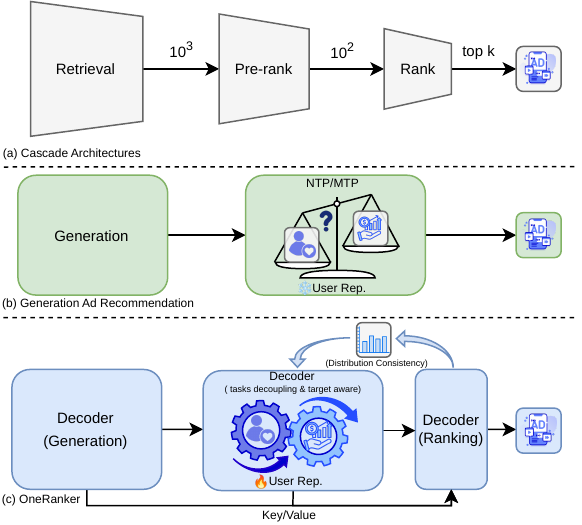}
% \caption{Comparison between Cascade Architectures, current generative recommendation systems and OneRanker.}
\caption{Comparison between existing methods and ours.}
\label{fig:figure-1}
\end{figure}

First, generative recommenders optimize multi-token prediction (MTP)~\cite{gloeckle2024better}, which generates candidate items from user behavior signals such as clicks and conversions. In advertising, however, the platform ultimately cares about business value, commonly measured by effective cost per mille (eCPM). Aligning these two objectives is difficult, and existing solutions face a dilemma. On one hand, \textit{stage decoupling} (the generate-then-rank pipeline) keeps the generator unaware of ranking objectives, so high-value candidates may be systematically filtered out at the generation stage, as shown in Figure~\ref{fig:figure-1}(a). On the other hand, \textit{single-stage fusion}---for example, directly injecting eCPM signals into the MTP heads---forces interest coverage and value optimization to compete within a shared representation space, inducing optimization tension, as shown in Figure~\ref{fig:figure-1}(b). 
Second, the inherent \textit{target-agnostic} limitation of generative models keeps user representations static during generation, unable to dynamically adapt to candidate item characteristics, constraining fine-grained user-item interaction capabilities. 
Third, the ``generate-then-rank'' architecture introduces two structural disconnections: representation inconsistency (heterogeneous representation spaces between generation and ranking) and error propagation (semantic drift in generation cannot be corrected during ranking), complicating joint optimization across stages. 
% These challenges collectively constitute the key bottleneck preventing generative advertising recommendation from transitioning from ``functional'' to ``excellent''.

To address these challenges, we propose \textbf{OneRanker}, whose core logic is shown in Figure~\ref{fig:figure-1}(c). 
First, as shown in Figure~\ref{fig:figure-2}, we introduce a value-aware multi-task decoupling architecture to achieve synergy between interest and value modeling. 
Specifically, we retain MTP's original multi-interest heads to maintain coverage breadth while introducing a dedicated value-aware head for business value learning. Both components share underlying user representations for knowledge transfer but decouple output spaces through independent task tokens, effectively alleviating optimization tension. 
Furthermore, we incorporate task ordering priors (e.g., impression $\rightarrow$ click $\rightarrow$ conversion $\rightarrow$ value) with causal mask, enabling high-level value tasks to dynamically absorb knowledge from low-level interest tasks, achieving progressive refinement from interest modeling to value optimization.

Secondly, to address the inherent target-agnostic limitation, we construct a coarse-to-fine collaborative target awareness mechanism. In Step 2 of Figure~\ref{fig:figure-2}, we introduce Fake Item Tokens obtained by K-means clustering of the entire item space to enable coarse-grained implicit awareness, allowing the generation process to dynamically perceive item semantic distributions. Simultaneously, we design a dual-channel representation mechanism (task semantic channel + target aware channel) that naturally fuses semantic matching and target awareness through inner product operations during retrieval, explicitly enhancing target sensitivity in generation. In Step 3 of Figure~\ref{fig:figure-2}, we introduce a ranking decoder that performs fine-grained target awareness through comprehensive cross-attention between candidate items and task tokens, enabling precise business value alignment.

Finally, to further enhance generation-ranking consistency and complete the unified framework, we propose dual-side consistency guarantees. On the input side, the ranking decoder constructs its Key/Value from both Step 1's original multi-interest representations and Step 2's refined representations, ensuring ranking decisions fully inherit generation process information. On the output side, we introduce a Distributional Consistency (DC) mechanism that propagates business value signals from Step 3's ranker back to Step 2's generator as soft labels. Through a joint loss function $\mathcal{L}_{\text{total}} = \mathcal{L}_{\text{MTP}} + \mathcal{L}_{\text{rank}} + \mathcal{L}_{\text{DC}}$, 
we achieve gradient-level collaborative optimization, enabling the generator to ``anticipate'' ranking preferences during training and fundamentally preventing objective drift between stages.

The main contributions of this work are summarized as follows:
\begin{itemize}
    \item We identify and alleviate the optimization tension between user interest modeling and business value optimization in generative advertising recommendation, proposing a value-aware multi-task decoupling architecture that achieves unified modeling of interest objectives and business value.
    \item We construct a coarse-to-fine collaborative target awareness mechanism: coarse-grained awareness via Fake Item Tokens in the generation stage, and fine-grained awareness via the ranking decoder for precise business value alignment.
    \item We design dual-side consistency guarantees through Key/Value fusion (input side) and joint loss optimization with DC (output side), enabling end-to-end collaborative optimization between generation and ranking.
    \item We have fully deployed OneRanker on Tencent's Weixin Channels advertising system. Large-scale online A/B tests show consistent gains on key business metrics (e.g., GMV-Normal +1.34\% in the initial 5\% traffic phase).
\end{itemize}

%————————————————————————————————%
% \begin{figure}[t]
% \centering
% \includegraphics[width=0.45\textwidth]{1.pdf}
% \caption{Comparison between previous methods and ours.}
% \label{fig:figure-1}
% \end{figure}
%————————————————————————————————%

\section{Related Work}
\subsection{Value-Awareness in Generative Models}

In recent years, the rise of generative models has shifted the recommender system paradigm from embedding-based matching to deep semantic understanding and generation. TIGER~\cite{tiger} employs RQ-VAE~\cite{rqvae} to produce quantized semantic identifiers, alleviating representation collapse in large-scale item spaces. GENRET~\cite{genret} co-optimizes identifier generation with retrieval objectives, enabling semantic IDs to dynamically capture complex interaction patterns. OneRec~\cite{onerec} proposes a generative recommendation model based on MoE~\cite{moe} and effectively improves sequence accuracy through DPO~\cite{dpo}. Furthermore, OneRec-Think~\cite{think} introduces Chain-of-Thought (CoT) reasoning~\cite{cot}, significantly enhancing decision-making logic and interpretability in complex scenarios.

Value alignment is particularly important in advertising recommendation. GPR~\cite{gpr} introduces value-aware fine-tuning that incorporates eCPM and action-value signals into multi-token prediction. More recently, GR4AD~\cite{xue2026generative} combines value-aware supervised learning with listwise reinforcement learning for large-scale generative advertising. These studies demonstrate the importance of introducing business-value signals into generative recommendation. OneRanker investigates a complementary architectural question: how to retain interest-oriented generation while modeling business value through separately parameterized task spaces and controlled information flow between them.
%This motivates our proposed Value-Aware Multi-Task Decoupled Architecture, which achieves native collaborative optimization of interest and value through output space decoupling.

\subsection{Target-Awareness Limitations in Generative Models}

Despite the rapid development of generative recommendation, existing generative methods still suffer from target-agnostic designs. In Transformer-based ~\cite{transformer} models~\cite{hstu,sasr,gsr}, user history is often compressed into a single embedding, which lacks fine-grained interaction with candidate items. Consequently, the user representation remains constant during the generation process and cannot be dynamically adjusted based on  candidate items. In advertising, where user intent is dynamic and influenced by multiple factors, this target-agnostic nature fundamentally limits the precision of identifying subtle user intentions from noisy behavioral signals.

Recent studies improve target awareness from different perspectives. GRank~\cite{grank} jointly trains a target-aware generator and a candidate-specific cross-attention ranker for large-scale retrieval. PROMISE~\cite{promise} introduces a process reward model that measures compatibility with the target candidate space, while GRAM~\cite{gram} controls autoregressive decoding using target attributes. OneRanker complements these approaches by combining learnable cluster anchors for coarse-grained target awareness with explicit candidate interaction in the ranking decoder.

\subsection{Joint Optimization of Generation and Ranking}

Traditional retrieval-ranking systems employ separately optimized candidate generation and ranking stages~\cite{huang2013learning,cheng2016wide,zhou2018deep,zhang2024wukong,zhu2025rankmixer}. Recent generative systems explore tighter coupling between generation and ranking from different perspectives. GRank~\cite{grank} jointly optimizes an MIPS-compatible generator and a candidate-specific cross-attention ranker for full-catalog retrieval. GoalRank~\cite{zhang2026goalrank} formulates ranking as generator-only $N$-to-$L$ list generation and applies group-relative listwise optimization. Other studies introduce ranking signals into generation through learning-to-rank or preference-alignment objectives~\cite{ltrgr,flame,rankgr,synergen}.

OneRanker focuses on the architectural integration of full-catalog generation and candidate-level ranking in industrial advertising recommendation. Steps 1 and 2 perform full-catalog generation, while Step 3 explicitly models candidate-level ordering within the same model. The stages exchange intermediate representations through Key/Value reuse, and a stop-gradient distribution-consistency objective transfers ranking preferences back to the generation side.

\section{Methods}

\subsection{Overall Architecture}
\label{subsec:overall_architecture}

\begin{figure*}[t]
\centering
\includegraphics[width=\textwidth]{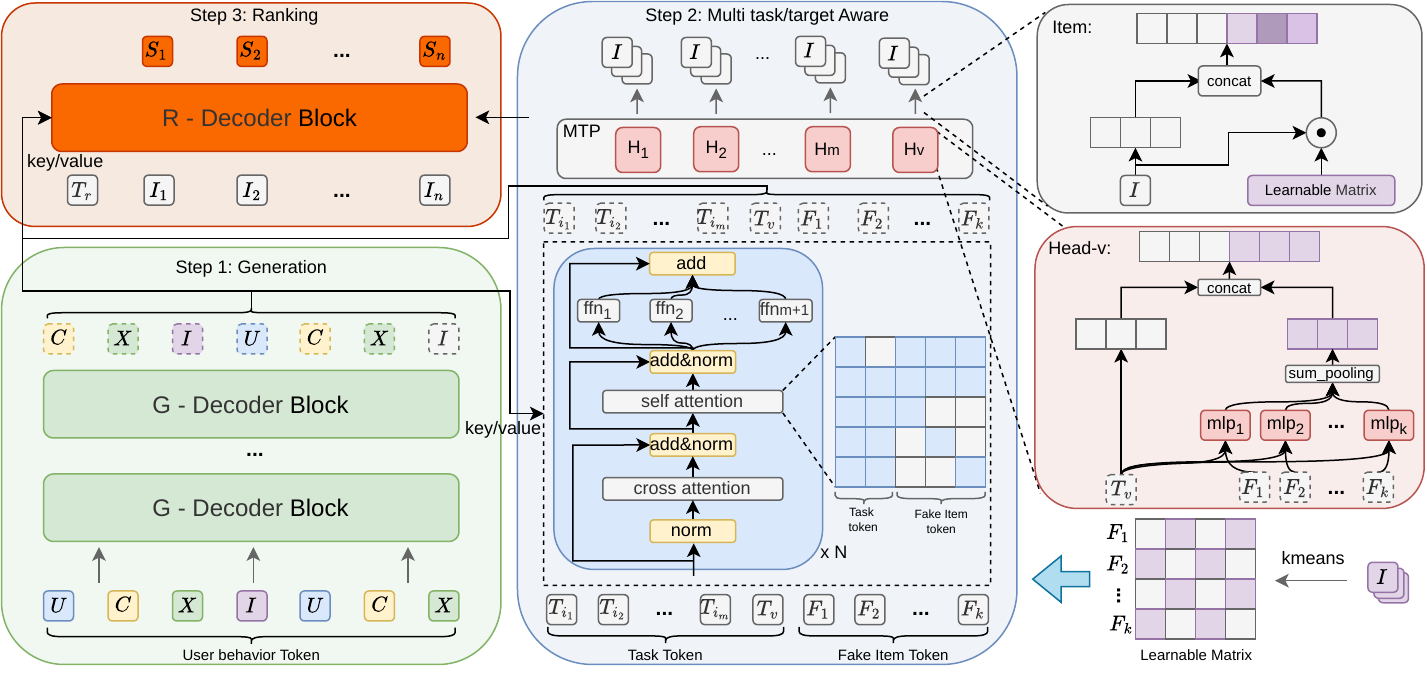}
\caption{The overall framework of OneRanker, consists of three stages: 
(i) the Generation stage which utilizes a generative backbone network pre-trained with multiple tokens based on user behavior sequences using heterogeneous tokens; (ii) the Multi-Task/Target-Aware stage which decouples interest and value objectives via task tokens and enables coarse-grained target awareness via fake item tokens; (iii) the Ranking stage which achieves fine-grained value alignment through a unified decoder with input-output consistency guarantees;}
\label{fig:figure-2}
\end{figure*}

We propose \textbf{OneRanker}, an end-to-end generative advertising recommendation framework, as illustrated in Figure~\ref{fig:figure-2}, which aims to systematically address three core challenges in generative advertising recommendation: (1) the objective misalignment between user interest modeling and business value optimization; (2) the inherent target awareness deficiency in the generation process; (3) the representation disconnection between generation and ranking stages. 
The framework consists of three logically progressive steps: \textbf{Step 1 (Generation)}, we adopt the same tokenization method as GPR \cite{gpr} to convert user behavior sequences into heterogeneous token streams (including user tokens, context tokens, content tokens, and item tokens), and perform multi-interest path generation based on the HSTU Decoder-only architecture. The autoregressive Multi-Token Prediction (MTP) mechanism can generate multiple complete semantic ID paths in parallel within a single forward pass, which has become a standard operation in industrial generative recommendation systems. \textbf{Step 2 (Multi-Task/Target-Aware)}, decouples interest and value objectives through task tokens and introduces fake item tokens to achieve coarse-grained target awareness. \textbf{Step 3 (Ranking)}, realizes business value alignment through fine-grained cross-attention mechanisms. The three steps are deeply coupled through Key/Value mechanisms.

%————————————————————————————————%
%————————————————————————————————%

\subsection{Generation Enhancement: Multi-Task/Target-Aware Enhancement}
\label{subsec:generation_enhancement}

This stage serves as an endogenous enhancement module within the generation pipeline, with its core objective being to enhance two critical capabilities of the MTP generation process: (1) \textbf{multi-task decoupling capability}---enabling different interest dimensions to perform specialized modeling for differentiated objectives, and separating interest modeling and business value optimization at the architectural level to alleviate their inherent optimization tension; (2) \textbf{coarse-grained target awareness capability}---mitigating the inherent ``target-agnostic'' limitation of generative models, enabling user representations to dynamically respond to candidate item characteristics. We achieve fine-grained calibration of multi-interest paths through the following four key designs, providing Step 3 with high-quality, target-aware input representations. This ensures that the final score alignment is built upon semantically rich and target-sensitive foundations, rather than ``post-hoc correction'' of low-quality generation results.

\subsubsection{Value-Aware Multi-Task Decoupling Architecture}
\label{subsubsec:value_aware_mtl}

We achieve architectural level decoupling of interest modeling and business value optimization through a unified task token sequence, rather than mixing objectives within a single-head representation space. We construct a learnable task token sequence $\mathbf{T} = [\mathbf{t}_{i_1}, \mathbf{t}_{i_2}, \dots, \mathbf{t}_{i_m}, \mathbf{t}_v]$, randomly initialized and optimized end-to-end during training. Specifically:
\begin{itemize}
    \item $\mathbf{t}_{i_1}, \dots, \mathbf{t}_{i_m}$ are \textbf{interest task tokens}, focusing on modeling differentiated interest dimensions and capturing multi-interest distributions through independent prediction heads. Each token serves an independent interest objective or business task or jointly forms a sequential task chain (e.g., impression $\rightarrow$ click $\rightarrow$ conversion).
    
    \item $\mathbf{t}_v$ is the \textbf{value-aware task token}, dedicated to learn business value (Final value).
\end{itemize}
All tokens share the underlying user representation but achieve task space decoupling through independent output heads, fundamentally alleviating the inherent tension between interest coverage and value optimization. The output at each task token position is routed to an independent prediction head, enabling explicit decoupling and parallel learning of multiple tasks. To model task dependencies, we propose the synergy of task ordering priors and causal mask: we introduce a task ordering prior---arranging tasks according to business sequence and cognitive difficulty (e.g., impression $\rightarrow$ click $\rightarrow$ conversion $\rightarrow$ business value)---while applying causal mask to ensure that subsequent tasks can access representations from preceding tasks, thereby enabling knowledge transfer and progressive refinement across tasks.

\subsubsection{Fake Item Token: Coarse-Grained Target Awareness}
\label{subsubsec:fake_item_token}

To mitigate the inherent ``target-agnostic'' limitation of generative models (i.e. the lack of interaction between user representations and candidate items), we introduce fake item tokens $\mathbf{F} = [\mathbf{f}_1, \dots, \mathbf{f}_k]$ as the core carrier for target awareness. 

These tokens are learnable parameters initialized as the k cluster center vectors obtained by performing K-means clustering over the entire item space, representing core anchors in the item semantic space. During training, they are optimized end-to-end, while simultaneously fitting item embeddings with gradients detached.
In the input sequence, we concatenate the Fake Item Tokens after the Task Tokens to form a complete Query sequence $\mathbf{Q} = [\mathbf{T}; \mathbf{F}]$; meanwhile, the output of Step 1 serves as the Key/Value for Cross-Attention. This design enables the model to dynamically perceive the item semantic distribution during the generation process, achieving coarse-grained implicit target awareness rather than relying solely on static user representations, thereby effectively alleviating the ``target-agnostic'' problem.

\subsubsection{Heterogeneous Attention Decoder}
\label{subsubsec:heterogeneous_attention}
We utilize the output of Step 1 as the Key/Value for the decoder, while the Query input comes from the constructed task tokens and fake item tokens $\mathbf{Q} = [\mathbf{T}; \mathbf{F}]$. Notably, we introduce two key improvements over the standard Transformer Decoder:

\begin{enumerate}
    \item \textbf{Cross-Attention Prioritization}: Traditional decoders first perform Self-Attention followed by Cross-Attention; we reverse this order. This design stems from a clear motivation: first aggregating and summarizing the user's multi-interest representations (Key/Value from Step 1) through Cross-Attention to obtain sufficient user-item interaction information, and then refining the aggregated Query (Task + Fake Item) through Self-Attention for internal consolidation and inter-task collaboration. Experiments show that this order better aligns with the cognitive logic of ``first understanding user intent, then refining task representations'', effectively enhancing representation quality.
    
    \item \textbf{Heterogeneous Mask Strategy}: To ensure reasonable information flow and prevent information leakage, we design a refined mask mechanism:
    \begin{itemize}
        \item \textbf{Among Task Tokens}: causal mask is applied to ensure temporal sequence modeling and dependency capture within the task sequence;
        \item \textbf{From Task Tokens to Fake Item Tokens}: each Task Token is allowed to access all Fake Items, enabling implicit target-aware interaction;
        \item \textbf{From Fake Item Tokens to Task Tokens}: each Fake Item is allowed to access all Task Tokens to fully capture multi-task semantics;
        \item \textbf{Among Fake Item Tokens}: mutual invisibility is enforced to avoid interference between different cluster centers, ensuring each Fake Item independently performs ``exploratory'' matching with user intent.
    \end{itemize}
\end{enumerate}

\subsubsection{Dual-Channel Representation Construction}
\label{subsubsec:dual_channel}

The outputs of the heterogeneous attention decoder (Task Tokens and Fake Item Tokens) are fed into the MTP module to construct refined multi-interest representations. The core innovation lies in the dual-channel representation fusion mechanism:

\begin{itemize}
    \item \textbf{Task Semantic Channel}: The decoder output at Task Token $\mathbf{T}_i$ is denoted by $\mathbf{e}_{\text{task}}^{(i)} \in \mathbb{R}^{d}$ and encodes the semantic representation for task $i$. In our implementation, task-token and item embeddings share the same dimension, i.e., $d=d_{\text{task}}=d_{\text{item}}$.
    
    \item \textbf{Target Aware Channel}: For each Task--Fake-Item pair $(\mathbf{T}_i,\mathbf{F}_j)$, their concatenated representation is mapped through a dedicated MLP to produce a $k$-dimensional contribution vector $\mathbf{s}^{(i)}_j \in \mathbb{R}^{k}$. Thus, each of the $k$ Fake Item Tokens produces one $k$-dimensional vector for task $i$. These vectors are aggregated via sum pooling to form the target-aware representation:
    \begin{equation}
        \mathbf{s}^{(i)}_{\text{target}} = \sum_{j=1}^{k} \mathbf{s}^{(i)}_j \in \mathbb{R}^{k}
    \end{equation}
    which explicitly models the user's preference distribution in the item semantic space under task $i$. We use $[\mathbf{s}^{(i)}_{\text{target}}]_l$ to denote its $l$-th component.
\end{itemize}

The final user representation for each MTP head $i$ is constructed by concatenating both channels:
\begin{equation}
    \mathbf{e}_{\text{user}}^{(i)} = \text{Concat}\left(\mathbf{e}_{\text{task}}^{(i)},\, \mathbf{s}_{\text{target}}^{(i)}\right) \in \mathbb{R}^{d + k}
\end{equation}

This design achieves explicit target awareness: $\mathbf{s}^{(i)}_{\text{target}}$ serves as a ``soft attention'' over the item semantic distribution, providing fine-grained target-aware signals for subsequent retrieval.

To achieve symmetric enhancement, the item-side representation similarly incorporates a target-aware channel: given an item embedding $\mathbf{e}_{\text{item}} \in \mathbb{R}^{d}$, we compute its cosine similarities with all $k$ cluster centers $\{\mathbf{f}_1, \dots, \mathbf{f}_k\}$ as $c_l = \text{cos\_sim}(\mathbf{e}_{\text{item}}, \mathbf{f}_l)$, and construct the symmetric enhanced item representation:

\begin{equation}
    \mathbf{e}_{\text{item}}^{\text{enhanced}} = \text{Concat}\left(\mathbf{e}_{\text{item}}, [c_1, c_2, \dots, c_k]\right) \in \mathbb{R}^{d + k}
\end{equation}

During retrieval, the relevance between user and item representations is computed via inner product:

\begin{equation}
    \text{score}(user, item) =
    \left(\mathbf{e}_{\text{user}}^{(i)}\right)^\top
    \mathbf{e}_{\text{item}}^{\text{enhanced}}
    =
    \left(\mathbf{e}_{\text{task}}^{(i)}\right)^\top
    \mathbf{e}_{\text{item}}
    + \sum_{l=1}^{k}
    [\mathbf{s}^{(i)}_{\text{target}}]_l \cdot c_l
    \label{eq:score_u_i}
\end{equation}

This inner product operation naturally fuses semantic matching with target-aware matching, endowing the retrieval process with explicit candidate item awareness and directly addressing the ``target-agnostic'' limitation of traditional generative models.

\subsection{Unified Ranking}
\label{subsec:unified_ranking}

\subsubsection{Fine-Grained Target Awareness}
\label{subsubsec:fine-grained}

This stage is responsible for transforming the refined representations from Step 2 into candidate-level ranking scores within OneRanker. We design a dedicated ranking decoder (R-Decoder Block), whose core innovation lies in internalizing the candidate-ranking process as a natural extension of the generation pipeline, rather than an external post-processing module. Specifically:

\textbf{Input Design}: The Query of the R-Decoder consists of two parts: (1) a dedicated ranking task token $\mathbf{T}_r \in \mathbb{R}^{d_{\text{task}}}$, serving as a semantic anchor for ranking intent; (2) $n$ candidate item tokens $\mathbf{I} = [\mathbf{i}_1, \dots, \mathbf{i}_n]$, representing the candidate set to be scored. During online serving, Steps 1 and 2 perform generation over the full item catalog and retain the top 500 candidates; Step 3 then orders these candidates before passing them to the separate downstream fine-ranking system. In our controlled offline evaluation, $n=30$ and the same fixed production candidate set is used by all compared methods rather than being independently regenerated by each model. Within OneRanker, these 30 candidates are processed through Steps 1 and 2 and are then passed unchanged to the R-Decoder for eCPM-oriented ordering. The Key/Value fuses the original multi-interest representations from Step 1 with the refined representations from Step 2, ensuring that ranking decisions simultaneously leverage the user's base interest distribution and the representations calibrated with target awareness.

\textbf{Mask Strategy}: Candidate item tokens $I$ are mutually invisible to each other (diagonal mask), avoiding interference among candidates and ensuring scoring independence for each candidate; All candidate item tokens $I$ can access the ranking task token ${T}_r$;

\textbf{Architecture and Output}: The R-Decoder adopts the same heterogeneous attention decoder structure as Step 2 (Cross-Attention prioritized followed by Self-Attention), but only use a single layer to complete ranking modeling. The output vector at each candidate item token position is directly mapped to a scalar score $s_i$ through a lightweight MLP. These scores $\mathbf{S} = [\mathbf{s}_1, \dots, \mathbf{s}_n]$ naturally reside in a unified semantic space and can be directly used for candidate ranking within OneRanker without additional normalization or calibration.

\subsubsection{Input-Side Consistency}
\label{subsubsec:input_consistency}

Input-side consistency connects the three stages through a \textbf{Key/Value pass-through mechanism}: the ranking decoder in Step 3 directly reuses the outputs of Step 1 and Step 2 as Key/Value, allowing generation-side representations to flow directly into candidate ranking.

This mechanism embodies dual semantic value:
\begin{itemize}
    \item \textbf{Step 1 $\rightarrow$ Step 3}: Transmits the user's base interest distribution, providing breadth coverage guarantees for ranking;
    \item \textbf{Step 2 $\rightarrow$ Step 3}: Transmits the intermediate representations used by the generation side to evaluate the candidate set, enabling the ranker to reuse and calibrate this information at a finer granularity.
\end{itemize}

This design effectively alleviates the \textbf{representation fragmentation} problem inherent in traditional ``external ranking'' approaches---the ranker no longer faces ``black-box'' generation results, but can instead perceive the internal states of the generation process. Therefore, while maintaining the multi-benefit coverage advantage of MTP, the system has achieved further accuracy improvements.

\subsubsection{Output-Side Consistency and Optimization Objectives}
\label{subsubsec:output_consistency}

Output-side consistency aligns the optimization objectives of generation and ranking through a \textbf{triadic joint loss function}, enabling end-to-end collaborative optimization. The total loss is defined as:
\begin{equation}
    \mathcal{L}_{\text{total}} = \mathcal{L}_{\text{MTP}} + \mathcal{L}_{\text{rank}} + \mathcal{L}_{\text{DC}}
\end{equation}
where $\mathcal{L}_{\text{MTP}}$ and $\mathcal{L}_{\text{rank}}$, respectively ensure the performance of the generation and ranking tasks, while $\mathcal{L}_{\text{DC}}$ is responsible for aligning the generation and ranking distributions. The components are defined as follows:

\textbf{Generation Loss ($\mathcal{L}_{\text{MTP}}$)}: We employ a negative log-likelihood loss to optimize multi-interest path generation in Step 2. For each task head $j$ and semantic ID level $l$, the loss is formulated as:
\begin{equation}
    \mathcal{L}_{\text{MTP}} = -\sum_{j=1}^{m} \sum_{l=1}^{L} \log P(\mathbf{e}_{\text{item}}^{\text{gt}} | \mathbf{e}_{\text{user}}^{(j)})
\end{equation}
where the conditional probability $P(\cdot)$ is constructed via softmax over the full semantic space $\mathcal{V}$ at that level:
\begin{equation}
\begin{split}
    P(\mathbf{e}_{\text{item}}^{\text{gt}} | \mathbf{e}_{\text{user}}^{(j)}) = \frac{\exp(s_{j,\text{gt}})}{\sum_{k \in \mathcal{V}} \exp(s_{j,k})}
\end{split}
\end{equation}
Here $s_{j,i} = \mathbf{e}_{\text{user}}^{(j)} \mathbf{e}_{\text{item}}^{(i)}$ represents the dot product score of user embedding and item embedding, the calculation details of which are shown in Equation~\ref{eq:score_u_i}. $\mathbf{e}_{\text{item}}^{\text{gt}}$ is the ground-truth semantic ID embedding for the next layer. 

Specifically, while interest heads optimize for click/conversion matching, the value-aware head employs value-weighted sampling strategies to optimize the generation distribution towards high-value items, ensuring value propensity of generated candidates.

\textbf{Ranking Loss ($\mathcal{L}_{\text{rank}}$)}: We adopt a pairwise BPR loss~\cite{rendle2012bpr} to optimize the relative ordering of candidates in Step 3:
\begin{equation}
    \mathcal{L}_{\text{rank}} = -\mathbb{E}_{(i,j) \sim \mathcal{D}} \left[ \mathbb{I}(y_i > y_j) \cdot \log \sigma(s_i - s_j) \right]
\end{equation}
where $(i,j)$ denotes an item pair, $y_i$ and $y_j$ are the \textbf{reference eCPM labels}, such that $y_i > y_j$ indicates that item $i$ has a higher reference eCPM than item $j$, and $s_i$ is the output of the Step 3 ranking decoder.

\textbf{Distributional Consistency Loss ($\mathcal{L}_{\text{DC}}$)}: Although the value-aware head optimizes true value propensity, its signals are sparse and lack perception of the ranker's \textbf{calibrated scoring space}. To bridge this gap, we use the ranker's candidate scores as dynamically updated teacher targets and align the generator's preference distribution with them. This objective is conceptually related to knowledge distillation~\cite{hinton2015distilling}, but the teacher and student are components of the same model and are trained concurrently through their respective objectives. For each optimization step, the target distribution is detached from the computation graph; therefore, $\mathcal{L}_{\text{DC}}$ updates the generator-side distribution but does not back-propagate through the ranker:
\begin{equation}
    \mathcal{L}_{\text{DC}} = -\sum_{i \in \mathcal{C}} \operatorname{sg}\!\left[p_i^{\text{target}}\right] \log \pi_\theta(i|\mathbf{u}, \mathcal{C})
    \label{eq:dc_loss}
\end{equation}
where $\mathcal{C}$ is the current candidate item set, $p_i^{\text{target}} = \text{softmax}(s_i / \tau)$ is the target probability distribution normalized from ranking scores, $\tau$ is the temperature parameter, and $\operatorname{sg}[\cdot]$ denotes stop-gradient. The ranker is optimized by $\mathcal{L}_{\text{rank}}$, while $\mathcal{L}_{\text{DC}}$ transfers its current preference distribution to the generator without creating a same-step feedback loop. Unlike the local pairwise optimization of $\mathcal{L}_{\text{rank}}$, $\mathcal{L}_{\text{DC}}$ constrains distributional consistency over the candidate set, enabling the generator to better anticipate the ranker's decision preferences.

Through dual-side consistency guarantee, the OneRanker framework builds a complete closed-loop optimization system: generation provides high-quality, target-aware candidates for ranking, while ranking supplies value-guided feedback signals for generation. 

\section{Experiments}
To evaluate the performance and robustness of our proposed model, we conducted extensive experiments to address the following research questions (RQs): 
\begin{itemize}[wide=0pt, leftmargin=*, labelindent=0pt, labelsep=0.5em, nosep]
    \item \textbf{RQ1:} How does OneRanker perform in comparison with current strong generative recommendation baselines in the industry?
    \item \textbf{RQ2:} How does the proposed key structural components contribute to the overall model performance?
    \item \textbf{RQ3:} How effective are the other dedicated designs of the proposed architecture ?
    \item \textbf{RQ4:} Does $\mathcal{L}_{\text{DC}}$ truly improve model consistency?
    \item \textbf{RQ5:} What is the empirical performance and real-world gain of OneRanker in an online industrial environment?
\end{itemize}

\begin{table*}[t!]
\setlength{\tabcolsep}{2pt}
\centering
% \fontsize{8.5pt}{10pt}
\caption{Comparison of Experimental Results on Key Metrics}
\label{tab:experimental_results}
\begin{tabular}{lcccccccccc}
\toprule
\textbf{Model/Method} & \textbf{HR@1} & \textbf{HR@3} & \textbf{HR@5} & \textbf{HR@10} & \textbf{HR@15} & \textbf{NDCG@1} & \textbf{NDCG@3} & \textbf{NDCG@5} & \textbf{NDCG@10} & \textbf{NDCG@15}\\
\midrule
HSTU & 0.1741 & 0.3508 & 0.4648 & 0.6604 & 0.7953  & 0.6742 & 0.6712 & 0.6763 & 0.7025 & 0.7396 \\
GPR & 0.1824 & 0.3703 & 0.4935 & 0.6957 & 0.8207 & 0.6823 & 0.6776 & 0.6818 & 0.7070 & 0.7445 \\
OneRanker & \textbf{0.2639} & \textbf{0.4959} & \textbf{0.6213} & \textbf{0.7945} & \textbf{0.8894} & \textbf{0.8102} & \textbf{0.7954} & \textbf{0.7904} & \textbf{0.7970} & \textbf{0.8206} \\
\bottomrule
\end{tabular}
\end{table*}

\subsection{Offline Experiments (RQ1)}
\subsubsection{Dataset Descriptions}
Following the experimental setup in GPR~\cite{gpr}, we conduct our evaluation on a large-scale real-world dataset collected from Tencent's Weixin Channels advertising platform, comprising hundreds of millions of anonymized users and a catalog of millions of items. The dataset comprises diverse user interactions across both advertising (impressions, clicks, and conversions) and organic content scenarios, ensuring robust representativeness and data scale. All historical behaviors are serialized in strict chronological order using a unified Heterogeneous Token Architecture (U/C/X/I), representing User, Content, Context, and Item tokens, respectively. To prevent information leakage, the data is split strictly by time: we use one year of interactions for training and the subsequent day for evaluation. Near-duplicate samples are filtered to reduce redundancy. In addition to sequential signals, the supervision includes reference eCPM labels and action types (impression, click, conversion) to reflect the value hierarchy. In our advertising setting, each reference eCPM label combines estimated user-response probabilities (pCTR and/or pCVR, depending on the optimization objective) with advertiser bid or conversion value, and therefore represents a joint response-value utility rather than a value-only signal. For each evaluation instance, a fixed set of 30 candidates is sampled from the online serving queue---i.e., real candidates that have survived the production retrieval and pre-ranking funnel and are passed to the fine-ranking stage. All methods (HSTU, GPR, OneRanker, and its ablation variants) receive exactly the same 30 candidates. Within OneRanker, this set is processed through Steps 1 and 2 before being passed unchanged to Step 3 for eCPM-oriented ranking. This controlled protocol isolates candidate scoring and ordering quality; it differs from online serving, where Steps 1 and 2 generate candidates from the full catalog and pass the top 500 to Step 3. Evaluating on these production-grade candidates, rather than on randomly sampled negatives, keeps the offline protocol close to the real serving distribution. All methods also share the same tokenizer and feature pipeline, so that the observed differences are not caused by discrepancies in the candidate universe or data processing.

\subsubsection{Baselines and Evaluation Metrics}
We benchmark OneRanker against two strong generative recommendation baselines:
\begin{itemize}[wide=0pt, leftmargin=*, labelindent=0pt, labelsep=0.5em, nosep]
    \item \textbf{HSTU~\cite{hstu}:} A trillion-parameter generative recommendation architecture developed by Meta, extensively deployed in large-scale industrial systems.
    \item \textbf{GPR~\cite{gpr}:} An end-to-end generative recommendation framework proposed by Tencent. Following the original paper, GPR not only optimizes interest via multi-token prediction but also injects eCPM and action-value signals through value-aware fine-tuning, making it a \textit{value-aligned} generative baseline rather than a purely interest-oriented model. GPR is currently the production baseline serving Weixin Channels advertising recommendation, which makes it the most relevant point of comparison for assessing the practical gains of OneRanker.
\end{itemize}
We adopt Hit Ratio (HR@K) and Normalized Discounted Cumulative Gain (NDCG@K) to evaluate candidate scoring and ranking quality on the fixed production candidate set. The eCPM-derived reference ordering is obtained by sorting the $n=30$ candidates by their reference eCPM labels in descending order. HR@K is one if the first candidate in this reference ordering appears in the model's predicted top-$K$ list, and zero otherwise; we report its average over all evaluation instances. For NDCG, a candidate at reference rank $r_i \in \{1,\ldots,n\}$ is assigned the linear relevance $\mathrm{rel}_i=n-r_i+1$. Given the model-predicted ordering $\pi$, we compute
\begin{equation}
    \mathrm{DCG@K}=\sum_{p=1}^{K}\frac{\mathrm{rel}_{\pi_p}}{\log_2(p+1)}, \qquad
    \mathrm{NDCG@K}=\frac{\mathrm{DCG@K}}{\mathrm{IDCG@K}},
\end{equation}
where $\mathrm{IDCG@K}$ is the DCG of the eCPM-derived reference ordering. Because each reference eCPM label combines user-response likelihood with advertiser value, these metrics evaluate candidate ordering with respect to their joint response-value utility. This offline protocol does not evaluate full-catalog candidate-generation recall.

\subsubsection{Implementation Details}
The embedding dimension for all token types is uniformly set to 128. In Step 1, the input U/C/X/I sequences are fixed to a maximum length of 2048 through truncation or padding, while the core G-Decoder module is configured with 4 Transformer layers. In Step 2, the number of interest-task tokens is set to 6 to align with the multi-task prediction heads. The value-aware task utilizes 2 specific task tokens whose outputs are aggregated via a bagging strategy. 32 Fake Item Tokens act as learnable parameters that characterize the semantic distribution of items; they are initialized as the cluster centers obtained by K-Means clustering over the entire item space and are subsequently updated end-to-end together with the model during training. The heterogeneous attention decoder consists of 2 layers. In Step 3, a single ranking-task token is used, and the R-Decoder depth is set to 1 layer for candidate-level calibration.

\subsubsection{Results}
Table~\ref{tab:experimental_results} summarizes the performance. While GPR consistently outperforms HSTU, OneRanker further improves over both baselines across all metrics. Specifically, HR@1 improves from 0.1824 to 0.2639 over GPR (a relative gain of 44.7\%), and similar gains are observed in ranking quality metrics; for example, NDCG@5 increases from 0.6818 to 0.7904, and NDCG@15 increases from 0.7445 to 0.8206.
These results suggest that our architecture, with its task-specific enhancements and value-aware ranking, can align candidate ordering more closely with the joint response-value objective represented by the reference eCPM labels. It is worth noting that GPR already optimizes business value through value-aware fine-tuning, i.e., by injecting value signals via objective weighting within a shared representation space---the single-stage value-injection strategy discussed in Section~\ref{sec:introduction}. The improvement of OneRanker therefore cannot be attributed to whether a business-value objective is introduced (both models optimize one), but rather to how value awareness and ranking are architecturally decoupled and then integrated within a single generative model. This is consistent with our central argument that it is architectural integration, rather than the mere addition of a value objective, that alleviates the optimization tension inherent to single-stage value injection.

\subsection{Ablation Studies (RQ2\&RQ3)}
\subsubsection{Ablation Studies on Key Structural Components (RQ2)}
To quantify the contributions of various components, we designed different model variants to address the aforementioned research questions. The details of each variant are as follows:
(1) \textbf{OneRanker}: The complete model. (2) Step 3 variants: \textbf{w/o DC Loss} (removes $\mathcal{L}_{\text{DC}}$), \textbf{w/o S2 Token Injection} (removes Step 2 tokens in K/V pairs), and \textbf{S3 Ranker Only} (removes both). (3) Step 2 variants: \textbf{OneRanker S2} (Step 2 only), \textbf{w/o Target} (removes Target-aware token), and \textbf{w/o Target \& MDA} (removes both Target-aware and Multi-task Decoupled Architecture).

\begin{table*}[htbp]
\setlength{\tabcolsep}{1pt}
\centering
% \footnotesize
% \fontsize{8.5pt}{10pt}
\caption{Ablation Studies Results of Different Model Variants}
\label{tab:ablation_study}
\begin{tabular}{lcccccccccc}
\toprule
\textbf{Model/Method} & \textbf{HR@1} & \textbf{HR@3} & \textbf{HR@5} & \textbf{HR@10} & \textbf{HR@15} & \textbf{NDCG@1} & \textbf{NDCG@3} & \textbf{NDCG@5} & \textbf{NDCG@10} & \textbf{NDCG@15}\\
\midrule
OneRanker (Full)                 & \textbf{0.2639}      & \textbf{0.4959}      & \textbf{0.6213}      & \textbf{0.7945}      & \textbf{0.8894}  & \textbf{0.8102} & \textbf{0.7954} & \textbf{0.7904} & \textbf{0.7970} & \textbf{0.8206}    \\
\quad w/o DC Loss          & 0.2629      & 0.4925      & 0.6173      & 0.7905      & 0.8867   & 0.8059     & 0.7912      & 0.7865      & 0.7936      & 0.8180   \\
\quad w/o S2 Token Injection & 0.2610      & 0.4912      & 0.6161      & 0.7897      & 0.8868  & 0.8047     & 0.7904      & 0.7858      & 0.7930      & 0.8172     \\
\quad with S3 ranker only    & 0.2601 & 0.4901 & 0.6157 & 0.7880 & 0.8852 & 0.8037 & 0.7894 & 0.7849 & 0.7924 & 0.8168 \\
OneRanker S2                     & 0.2214 & 0.4257 & 0.5448 & 0.7291 & 0.8435 & 0.7566 & 0.7460 & 0.7440 & 0.7579 & 0.7876\\
\quad w/o Target          & 0.2080 & 0.4030 & 0.5203 & 0.7066 & 0.8276 & 0.7375 & 0.7286 & 0.7285 & 0.7465 & 0.7786 \\
\quad w/o Target \& MDA   & 0.2035 & 0.3922 & 0.5066 & 0.6949 & 0.8212 & 0.7360 & 0.7273 & 0.7275 & 0.7453 & 0.7776 \\
% \quad w/o Value           & 0.0660 & 0.1711 & 0.2611 & 0.4537 & 0.6221 \\
\bottomrule
\end{tabular}
\end{table*}

Experimental results in Table \ref{tab:ablation_study} demonstrate that each core module significantly contributes to performance. We focus on HR@5 and NDCG@5 to evaluate both hit accuracy and ranking quality.

In Step 2 , compared to the Base (OneRanker S2), removing the Target information (\textit{w/o Target}) causes a notable decline, with HR@5 dropping by 4.5\% and NDCG@5 decreasing by 2.1\%, which strongly proves the necessity of the Fake-Item-Token mechanism, prompting the model to evolve from static user representations to dynamic-aware item semantic distribution representations, thereby capturing users' real-time intentions more accurately. 
Furthermore, the simultaneous removal of Target information and the Multi-task Decoupled Architecture (w/o Target \& MDA) leads to a further decline in HR@5 and NDCG@5, underscoring the critical role of the decoupled architecture in mitigating optimization tension within the shared representation space, effectively preventing interference between business value modeling and user interest modeling.

For the Step 3 ablations, the S3 Ranker Only variant, which removes both Step 2 token injection and $\mathcal{L}_{\text{DC}}$, serves as the controlled reference. Retaining Step 2 token injection while removing $\mathcal{L}_{\text{DC}}$ (w/o DC Loss) yields an HR@5 of 0.6173 and an NDCG@5 of 0.7865. Conversely, retaining $\mathcal{L}_{\text{DC}}$ while removing Step 2 token injection (w/o S2 Token Injection) yields 0.6161 and 0.7858, respectively. These two variants isolate the contributions of input-side information reuse and output-side distribution alignment as parallel mechanisms.
Combining both mechanisms produces the full OneRanker, which achieves an HR@5 of 0.6213 and an NDCG@5 of 0.7904, corresponding to relative improvements of 0.91\% and 0.70\% over S3 Ranker Only. The larger gain of the combined configuration suggests that the two consistency mechanisms are complementary.

\begin{table}[htbp]
\setlength{\tabcolsep}{1pt}
\centering
% \footnotesize
\fontsize{8.5pt}{10pt}
\caption{Ablation Studies on Dedicated Designs in Step 2}
\label{tab:ablation_step2}
\begin{tabular}{lccccc}
\toprule
\textbf{Model/Method} & \textbf{HR@1} & \textbf{HR@3} & \textbf{HR@5} & \textbf{HR@10} & \textbf{HR@15} \\
\midrule
OneRanker S2 (Baseline) & 0.2214 & 0.4257 & 0.5448 & 0.7291 & 0.8435 \\
\quad w/o CA-Pri & 0.2117 & 0.4100 & 0.5277 & 0.7148 & 0.8337 \\
\quad w/o H-Mask & 0.2149 & 0.4151 & 0.5335 & 0.7183 & 0.8368 \\
\bottomrule
\end{tabular}
\end{table}

\subsubsection{Ablation Studies on Dedicated Designs (RQ3)}
We further investigate the effectiveness of several dedicated architectural design choices within Step 2.
Specifically, we implement a Cross-Attention Prioritization (CA-Pri) Mechanism, which allows each task-specific token to sufficiently aggregate user information prior to refining task representations. Additionally, a Heterogeneous Mask Strategy is designed to regulate the information flow while preventing information leakage between disparate task tokens. 
As shown in Table~\ref{tab:ablation_step2}, removing either the Cross-Attention Prioritization (CA-Pri) Mechanism or the heterogeneous masking (H-Mask) strategy from the Step 2 baseline leads to a significant performance degradation across all metrics. These results underscore the necessity of our refined attention mechanism and mask strategy in achieving accurate recommendation.

\subsection{Effectiveness Analysis of $\mathcal{L}_{\text{DC}}$ (RQ4)}
\begin{figure}[h]
     \centering
     \begin{subfigure}[b]{0.23\textwidth}
         \centering
         \includegraphics[width=\textwidth]{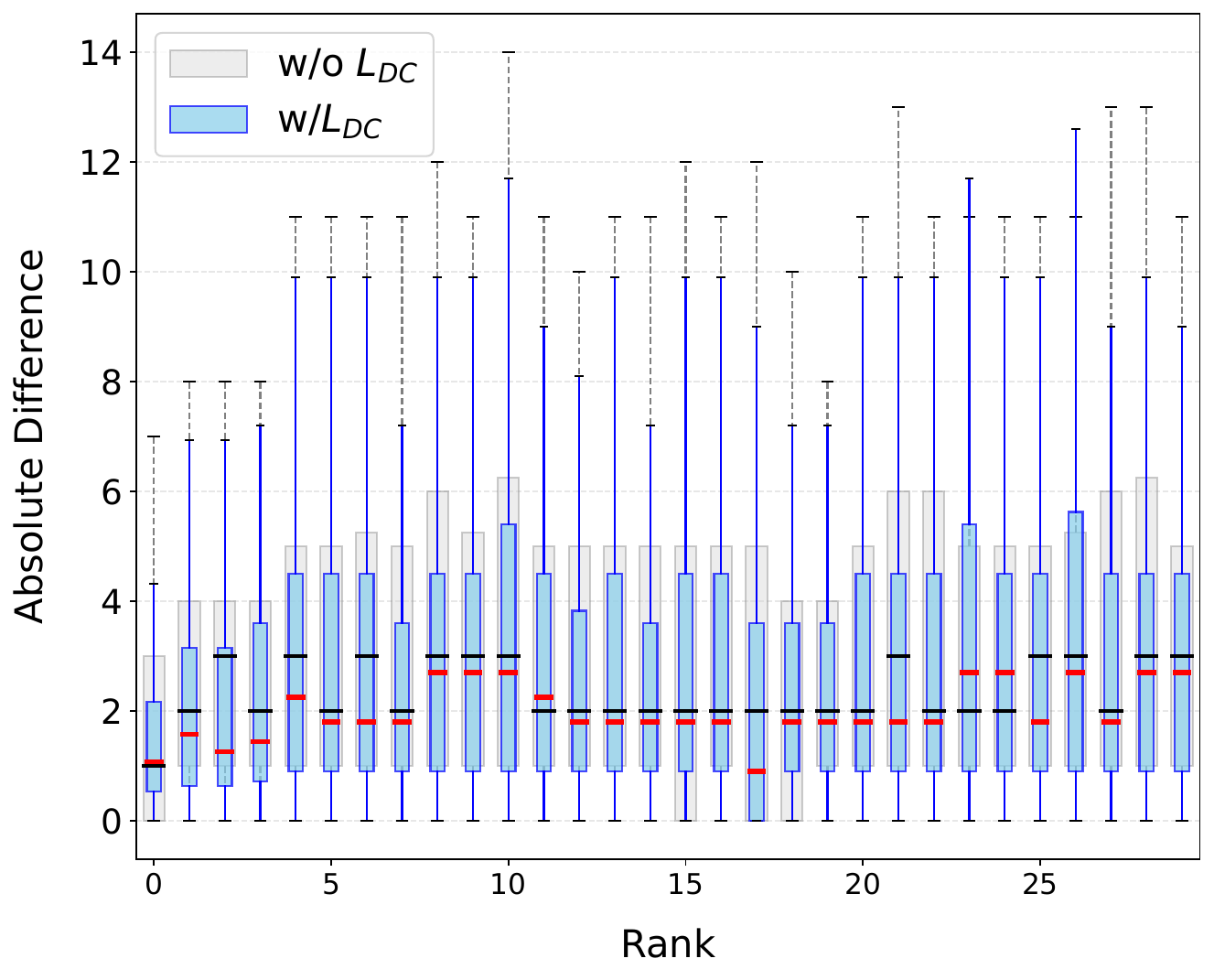}
         \caption{Absolute Rank Difference}
         \label{fig:rank_deviation}
     \end{subfigure}
     \hfill
     \begin{subfigure}[b]{0.235\textwidth}
         \centering
         \includegraphics[width=\textwidth]{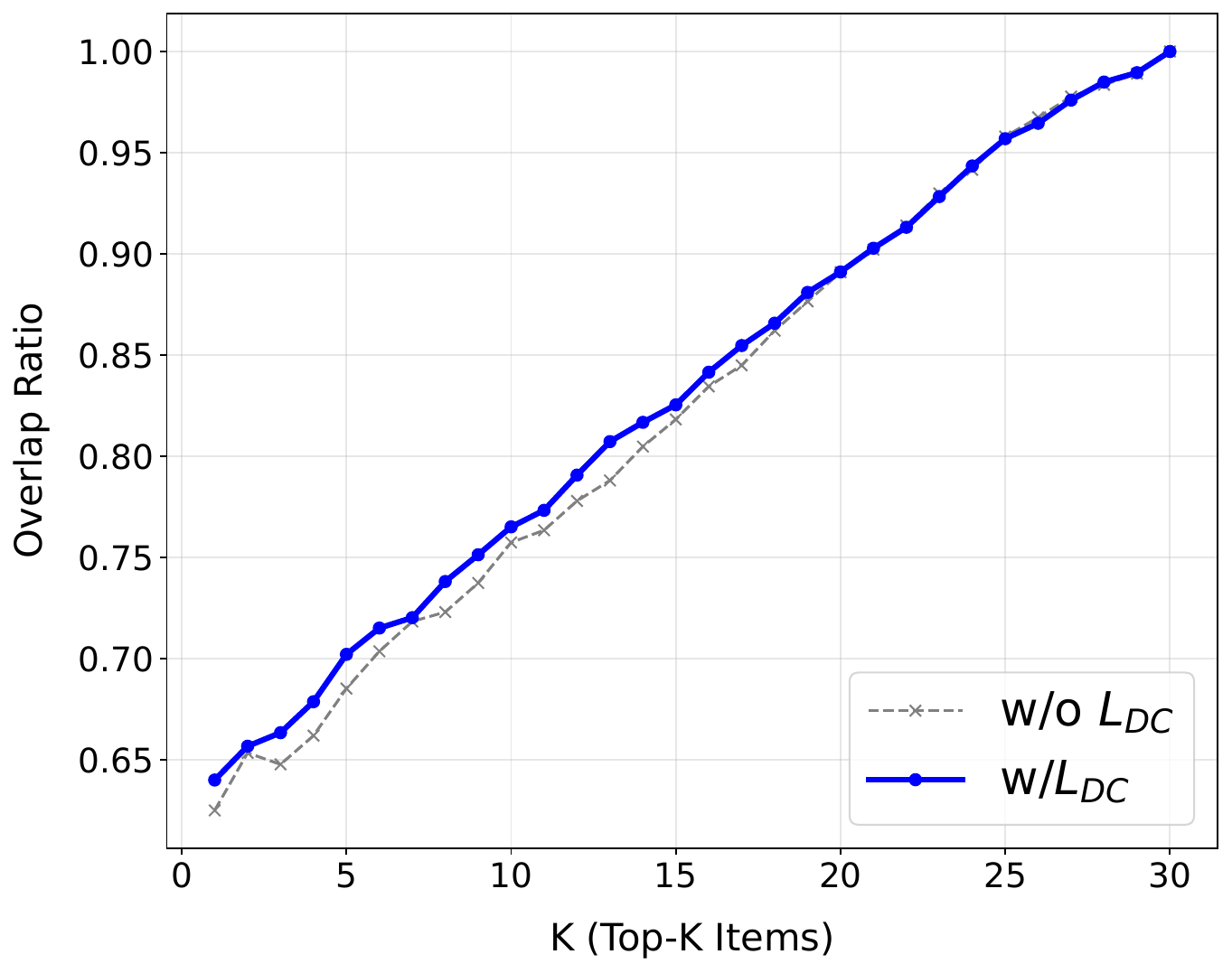}
         \caption{Top-$K$ Item Overlap Ratio}
         \label{fig:overlap_curve}
     \end{subfigure}
     
     \caption{Effectiveness of $\mathcal{L}_{\text{DC}}$: (a) Absolute rank difference of items between Step 2 and Step 3 at each rank for models with and without $\mathcal{L}_{\text{DC}}$; (b) Top-$K$ items overlap ratio between Step 2 and Step 3 with varying $K$ for the two models.}
     \label{fig:alignment_results}
\end{figure}

To quantitatively evaluate the consistency between models with and without the Distributional Consistency Loss $\mathcal{L}_{\text{DC}}$ , we employ rank deviation boxplots and Top-K overlap curves across 30 candidate items, as is shown in Figure~\ref{fig:alignment_results}.
Figure~\ref{fig:rank_deviation} characterizes the consistency by measuring the absolute rank differences of Step 2 and 3: the central box represents the interquartile range (IQR) defined by the first (Q1) and third (Q3) quartiles, while the internal horizontal line denotes the median and whiskers represent the minimum and maximum respectively. Figure~\ref{fig:overlap_curve} measures the alignment degree between the two steps when filtering the top $K$ items.

Experimental results demonstrate that incorporating the Distributional Consistency (DC) loss compresses the IQR (blue boxes), as evidenced by the markedly reduced box height compared to the baseline in Figure~\ref{fig:rank_deviation}. This contraction reflects a high degree of decisional consistency of the two steps and enhanced stability. 
Correspondingly, the Top-K overlap curve in Figure~\ref{fig:overlap_curve} validates this distributional optimization: the optimized blue curve exhibits a higher starting point at $K=1$ and maintains a consistent lead until $K$ approaches the total candidate count. This confirms that the reduction in individual rank variance observed in the Figure~\ref{fig:rank_deviation} directly translates into stronger consistency effect, establishing a more robust multi-stage collaborative framework.

\subsection{Online A/B Test (RQ5)}

We evaluated and deployed OneRanker on Weixin Channels, a production advertising platform with hundreds of millions of active users and tens of millions of dynamic ads. The primary online metrics include GMV, GMV-Normal (ads optimized for clicks or conversions, constituting the majority of total GMV), and Costs. Among these, GMV-Normal serves as the core value-oriented Key Performance Indicator (KPI), as it most directly reflects the business returns from the value-optimized traffic that OneRanker targets, while GMV and Costs are reported alongside it for completeness.

Our online baseline is the production cascaded pipeline, in which GPR serves as the candidate generation (retrieval) stage and feeds a separate downstream pre-ranking module, followed by a shared fine-ranking system. In OneRanker, Steps 1 and 2 perform full-catalog generation and retain the top 500 candidates, which are then ordered by the Step 3 ranking decoder before being passed to the unchanged downstream fine-ranking system. OneRanker therefore replaces the retrieval and pre-ranking stages of the production pipeline with a single unified model, while the downstream fine-ranking stage is kept unchanged. This constitutes a controlled comparison against the strong cascaded system actually running in production: because only the retrieval and pre-ranking stages differ, the A/B outcome can be reasonably attributed to consolidating these two separate stages into one integrated model. After rigorous A/B testing, OneRanker achieved stable and statistically significant improvements on the core value-oriented KPI compared to the previous version. The results are summarized in Table~\ref{tab:online_results}.
We report these online results transparently. GMV-Normal, the primary value-oriented KPI, improves significantly and consistently across all three phases. Although total GMV gains are not always statistically significant and Costs rise, guardrail metrics such as CTR and CVR also improve significantly, indicating that the additional spend reflects higher-quality user engagement rather than wasteful exposure. Based on these results, OneRanker was fully rolled out to 100\% traffic for these stages in the Weixin Channels advertising system. We note that the measured lift narrows as the treatment traffic grows, which is primarily attributable to traffic coverage effects rather than a decay in model effectiveness. To obtain a cleaner estimate under the fully deployed steady state, we conducted a 5\% reverse (holdback) experiment immediately after the full rollout, which measured a +2\% improvement in GMV-Normal, confirming that OneRanker delivers sustained gains at full production scale.

\begin{table}[htbp]
\centering
\caption{Online A/B Test Results of OneRanker. Confidence intervals (CI) are calculated with 0.05 significance level.}
\label{tab:online_results}
\small % 适当缩小字号
\setlength{\tabcolsep}{6pt} % 调整列间距，根据宽度可增减
\begin{tabular}{lccc}
\toprule
\textbf{Traffic Phase} & \textbf{GMV (CI)} & \textbf{GMV-Normal (CI)} & \textbf{Costs (CI)} \\
\midrule
5\% Phase & +0.4067\% & +1.3427\% & +0.7190\% \\
    & \scriptsize{[-0.8320\%, 1.6454\%]} & \scriptsize{[0.1637\%, 2.5216\%]} & \scriptsize{[0.4632\%, 0.9747\%]} \\
\addlinespace[0.5em] % 增加行间距，区分不同的 Phase
20\% Phase & +0.7796\% & +0.6446\% & +1.1462\% \\
     & \scriptsize{[0.2849\%, 1.2743\%]} & \scriptsize{[0.2139\%, 1.0754\%]} & \scriptsize{[0.9958\%, 1.2965\%]} \\
\addlinespace[0.5em]
80\% Phase & +0.0843\% & +0.3500\% & +0.4493\% \\
     & \scriptsize{[-0.4531\%, 0.6217\%]} & \scriptsize{[0.0188\%, 0.6812\%]} & \scriptsize{[0.3261\%, 0.5725\%]} \\
\bottomrule
\end{tabular}
\end{table}

\section{Conclusion}
\label{sec:conclusion}

This paper presents OneRanker, systematically addressing three core challenges in generative advertising recommendation: the misalignment between interest and value objectives, the target-agnostic limitation of generation processes, and the architectural disconnection between generation and ranking. By leveraging value-aware multi-task decoupling, coarse-to-fine collaborative target awareness, and input-output dual-side consistency guarantees, OneRanker achieves architectural-level integration and end-to-end collaborative optimization. A full-scale deployment on Tencent's Weixin Channels advertising system validates its effectiveness, yielding a +1.34\% improvement in GMV-Normal in the initial rollout phase. Rather than claiming to replace cascaded systems outright, we present OneRanker as a practical step toward tighter integration of generation and ranking, and we discuss the trade-offs (e.g., increased cost and added model complexity) observed during deployment, accumulating practical experience for applying generative techniques in complex industrial scenarios.

% The acknowledgments section is defined using the "acks" environment
% (and NOT an unnumbered section). This ensures the proper
% identification of the section in the article metadata, and the
% consistent spelling of the heading.
% \begin{acks}
% \end{acks}

\bibliographystyle{ACM-Reference-Format}
\bibliography{sample-base}

\appendix

\end{document}